\magnification=1200
\tolerance=500
\hsize 15.0truecm\hoffset 0.7truecm
\baselineskip=15pt                       

\parindent 0.9truecm
\outer\def\beginsection#1\par{\medbreak\bigskip
      \message{#1}\leftline{\bf#1}\nobreak\medskip\vskip-\parskip
      \indent}

\def\ref#1{[#1]}  

\ifx\usepostscript\undefinedcs  
\else
\fi
\def\ee{$e^+e^-$}               



\def\NF{{\cal N}_{\kern -1.9pt f}}     
\def\NC{{\cal N}_{\kern -1.7pt c}}     

\def\pt{{p\kern -.2pt\lower 4pt\hbox{\fivei T}}}    

\def\pl{{p\kern -.2pt\lower 4pt\hbox{\fivei L}}}

\def\nostrocostrutto#1\over#2{\mathrel{\mathop{\kern 0pt \rlap 
  {\raise.2ex\hbox{$#1$}}}
  \lower.9ex\hbox{\kern-.190em $#2$}}}
\def\lsim{\nostrocostrutto < \over \sim}   

\pageno=0
\begingroup
\nopagenumbers

\null
\vskip -1ex
\baselineskip=14pt
\rightline{DFTT 7/94}
\rightline{21 March, 1994}

\vskip 4.truecm
\centerline{\bf CLAN PROPERTIES IN PARTON SHOWERS}
\vskip 1truecm   
\centerline {R.\ Ugoccioni, \quad A.\ Giovannini, \quad S.\ Lupia}
\smallskip
\centerline {\it Dipartimento di Fisica Teorica, Universit\`a di Torino}
\centerline {\it and INFN, Sezione di Torino}
\centerline {\it via P. Giuria 1, 10125 Torino, Italy}
\vskip 1.4truecm  

\parindent 0cm
\footnote{}{E-mail addresses: giovannini@to.infn.it, lupia@to.infn.it,
ugoccioni@to.infn.it}
\footnote{}{\it Work supported in part by M.U.R.S.T.\ 
(Italy) under Grant 1993}
\parindent 1cm

\vfil
\midinsert
\parindent 1.5truecm
\narrower
\centerline{ {\bf ABSTRACT} }
\smallskip
\noindent By considering clans as genuine 
elementary subprocesses, {\it i.e.}, intermediate parton sources 
in the Simplified Parton Shower  model, a generalized version of this 
model is defined. It predicts analytically clan properties at 
parton level in agreement with the general trends observed 
experimentally at hadronic level and in Monte Carlo simulations both 
at partonic and hadronic level. In particular the model shows 
a linear rising in rapidity 
of the average number of clans at fixed energy of the initial parton and 
its subsequent bending for  rapidity intervals at the border of 
phase space, and approximate energy independence of the average number 
of clans in fixed rapidity intervals.
The energy independence becomes stricter by properly normalizing the 
average number of clans.

\endinsert

\vfil
\eject
\endgroup

\outer\def\subsection#1\par{\medbreak\medskip
      \message{#1}\leftline{\it#1}\nobreak\medskip\vskip-\parskip
      \indent}
\outer\def\romansubsection#1\par{\medbreak\medskip
      \message{#1}\leftline{#1}\nobreak\medskip\vskip-\parskip
      \indent}
\def\psigma{{p\kern -.2pt\lower 4pt\hbox{$\scriptstyle\Sigma$}}}   
\def\rsigma{{r\kern -.2pt\lower 4pt\hbox{$\scriptstyle\Sigma$}}}   
\def\P{{\cal P}(Q_0Q_1|Q)}

\def\lw{\log W}
\def\la{\log 2}
\def\yc{y_c}
\def\spac{\noalign{\vskip6pt}}

\null

\beginsection Introduction

Clan concept was introduced in \ref{1} in order to interpret 
the wide occurrence of NB  
regularity in terms of a two step cascade process. 
Clans are defined, 
as well known, as group of particles of common ancestor; they are 
independently produced and each clan contains at least one particle. 
With these 
assumptions the average number of clans is distributed according to a 
Poisson distribution. Negative Binomial (NB) 
behavior is obtained by requiring in addition 
that particles inside an average clan have a logarithmic distribution 
which results from the convolution of truncated geometric 
distributions for particles within single clans\ref{2}. 

Clan structure analysis in terms of the average number of 
clans, $\bar N$, and of the average number of particles per clan,  
$\bar n_c$, at final hadron 
level has given striking and unexpected results which are still 
puzzling. In particular it has been 
shown that the density of clans per unit of rapidity in symmetric 
rapidity intervals, although different in \ee\ annihilation\ref{3} 
from deep inelastic scattering\ref{4} 
and hadron-hadron collisions\ref{5}, is approximately 
independent of the c.m.\ energy of the collision; furthermore 
the average number of clans at 
fixed c.m.\ energy grows initially linearly with the width of the 
rapidity interval, then 
a characteristic bending is seen at the border of phase space.

It should be pointed out that these regularities are satisfied with a 
higher degree of accuracy at final parton level than at final hadron 
level as it was seen 
by applying clan structure analysis to the final 
parton Multiplicity Distributions (MD's) in symmetric rapidity 
intervals reconstructed  {\it via} Generalized Local Parton Hadron 
Duality (GLPHD) hadronization prescription\ref{6} 
from the corresponding MD's at final hadron level\ref{7}.
The conclusion was that a complex phenomenon like Multiparticle 
Production might very well be at parton level more elementary than at 
final hadron level.

The natural question at this point was the following: is simplicity 
maintained at an even more elementary level of investigation, {\it i.e.},
 at single jet level? Single quark- and gluon-jets 
can be isolated as well known in experiments 
and Monte Carlo  simulations 
using a convenient jet-finding algorithm.
The use of these algorithms, although still under discussion, 
should be considered at 
present the only way which we have to localize single jet properties 
from many jet events samples.
In this connection, it was interesting to find in Monte Carlo 
simulations that the 
mentioned regularities for clans are valid with a high degree of 
precision for final hadrons in single jets originated by an initial 
quark and gluon\ref{8}.
The accuracy is again improved by going {\it via} GLPHD to final parton 
level.

In parallel, the theoretical study of single parton shower was progressing in the 
literature\ref{9,10,11,12}. A contribution along this line of thought is 
the Simplified Parton Shower (SPS) model\ref{13}, which 
describes final parton MD's in a single jet originated by an initial 
parton assuming that partons are controlled by the dominant interaction  
among gluons (gluon self-interaction) in a 
correct kinematical framework.
The model was solved numerically and 
results confirm the relevance of clan structure analysis for discussing 
final parton MD's properties.

New perspectives on the theoretical basis of clan concept were opened 
later on by the discovery of its link with void probability 
in rapidity space, {\it i.e.}, the probability to detect no 
particles in a given domain of phase space,  and the  
related hierarchical structure of the corresponding correlation 
functions\ref{14}: the 
relevance of clan concept emerged fully within the class of Infinitely 
Divisible Distributions which reduce to NB MD in a particular 
case, {\it i.e.}, the clan concept was found to be more general than NB 
regularity.

All these undeniable successes  notwithstanding, our search failed in 
explaining the independence of the average 
number of clans on c.m.\ energy in a fixed rapidity interval, its 
linear dependence on rapidity at fixed c.m.\ energy and its subsequent 
bending for larger rapidity intervals.
To our knowledge, it should be added, these facts 
are neither explained nor predicted in single parton showers (jets) 
by any of the existing models in the field.

In order to answer these challenging questions at parton level,  
we decided to implement clan 
concept in the SPS model\ref{13}. 
This generalized version of SPS, which we call 
Generalized Simplified Parton Shower  (GSPS) model goes as follows.
We assume that the number 
of clans in each event coincides with the number of steps of the 
Markov cascade originated by an initial parton, {\it i.e.}, at each step 
in the evolution of a single jet independent intermediate shower sources, 
the clans, are emitted with degrading virtualities.
Consequently, the number of clan 
ancestors is distributed according to a shifted Poisson 
distribution.
By assuming local fluctuations in virtuality of the initial parton at 
the origin of the cascade and of clan ancestors, 
we calculate {\it analytically} the energy and rapidity 
dependence of the average number of clans.
It is shown that clan properties at partonic single jet level 
obtained from experimental data  and from Monte Carlo 
simulations via GLPHD are correctly reproduced by our model. 
A new striking regularity for the average number 
of clans is also discovered. 


In section {\bf II} we review the structure of the SPS model for single jets; 
in section {\bf III} we present its generalized version and calculate clan 
MD's both in full phase space and in symmetric rapidity intervals. 
Conclusions are drawn at the end.

\beginsection II. Summary and critical reading of SPS model for single jets.

In this section we examine the main features of SPS model which are 
relevant for our scope and which we developed in 
\ref{13}. Our aim was to obtain the virtuality 
evolution and the rapidity structure of a single parton shower 
by assuming essentials of QCD 
and the validity at each step of the cascading process 
of the energy-momentum conservation law. 
Inspired by the criterium of maximum simplicity, 
essential features of QCD which we decided to incorporate in the model 
are the dominance of gluon self-interaction as described by a kernel of 
the form suggested in \ref{15} and 
a Sudakov form factor term in order to normalize the 
elementary virtuality splitting function.
Virtuality evolution in a single jet (parton shower) was then 
described as follows.

We consider an initial parton of maximum allowed virtuality $W$ which splits at virtuality 
$Q$ into  two partons of virtuality $Q_0$ and $Q_1$. We require $Q \ge Q_0 + Q_1$ 
and $Q_0, Q_1  \ge 1$ GeV. We define the probability for a parton of virtuality $W$ to split 
at $Q$, $p(Q \vert W)$, which is normalized by a Sudakov form factor.
The probability for an ancestor parton 
of maximum allowed virtuality $W$ to generate $n$ final partons, 
 $P_n(W)$, and the probability for a parton which splits 
at virtuality $Q$ to 
generate $n$ final partons, $R_n(Q)$, with $R_n(Q) = \delta_{n1}$ 
for $Q < 2$ GeV, 
lead to the corresponding generating functions:
$$
f(z,W) = \sum_{n=1}^{\infty} P_n(W) z^{n-1}
\eqno(1)
$$
$$
g(z,Q) = \sum_{n=2}^{\infty} R_n(Q) z^{n-2} 
\eqno(2)
$$
The two generating functions are linked by
$$
f(z,W) = \int_1^2 p(Q \vert W) dQ + \int_2^W p(Q \vert W) z g(z,Q) dQ 
\eqno(3)
$$
The joint probability density $\P$ for a parton of virtuality $Q$ to 
split into two partons of virtuality $Q_0$ and $Q_1$ is defined by  
$$
\P = p(Q_0 \vert Q) p(Q_1 \vert Q) K(Q) 
\eqno(4)
$$
where $K(Q)$ is a normalization factor.

The dynamical content of the model in {\it virtuality} 
is contained in the following 
equation
$$
R_n(Q) = \sum_{n'=1}^{n-1} \int_1^{\infty} dQ_0 \int_1^{\infty} dQ_1 \P R_{n-n'}(Q_0) 
R_{n'}(Q_1) \theta(Q-Q_0-Q_1) 
\eqno(5) 
$$
which gives the probability for a parton which splits 
at virtuality $Q$ to 
generate $n$ final partons  in terms of 
the joint probability density for a parton of virtuality $Q$ to 
split into two partons of virtuality $Q_0$ and $Q_1$.

\noindent
Eq. (5) can be reformulated for the corresponding generating function 
by dividing the domain of integration in three subdomains; one obtains 
$$\eqalign{
g(z,Q) =&~ \int_1^2 dQ_0 \int_1^2 dQ_1 \P \theta (Q-Q_0-Q_1) + \cr
&~2 \int_1^2 dQ_0 \int_2^{\infty} dQ_1 \P z g(z,Q_1) \theta (Q-Q_0-Q_1) + \cr
&~\int_2^{\infty} dQ_0 \int_2^{\infty} dQ_1 \P zg(z,Q_0) zg(z,Q_1) 
\theta (Q-Q_0-Q_1)\cr}
\eqno(6)
$$
The above three subdomains correspond to the possible different 
situations in which the two generated partons can be found {\it i.e.}, 
no one or only one or both split. 
This general scheme is valid for any splitting function $p(Q \vert W)$. 
In case $p(Q \vert W)$ 
is factorizable in terms of its variables $Q$ and $W$, {\it i.e.}, 
$$
p(Q \vert W) = p_0(Q) C(W), \quad 
C(W) = \left[ \int_1^W dQ' p_0(Q') \right]^{-1} 
\eqno(7)
$$
eq.~(3) simplifies into the differential equation
$$
{\partial f(z,W) \over \partial W} = p(W \vert W) [z g(z,W) - f(z,W)]
\eqno(8)
$$
For numerical simulations we choose:
$$
p(Q \vert W)dQ = {A \over Q} {(\log Q)^{A-1} \over (\log W)^A} dQ =d\left( {\log Q \over \log W} \right)^A
\eqno(9)
$$
$A$ being the only free parameter of the model. 
This form of $p(Q|W)$ was motivated in \ref{13} by our request of 
simplicity in the structure of the model. Notice that eq. (9) 
corresponds to the virtuality dependence of the standard QCD kernel.

For describing the {\it rapidity} structure of the model, we proposed 
in \ref{13} 
to use the singular part of the QCD kernel controlling gluon branching:
$$
p(y_0|Q_0Q_1Qy) \propto 
P(z_0)dz_0 \propto {dz_0 \over z_0 ( 1 - z_0)}
\eqno(10) 
$$
Here $z_0$ is the energy fraction carried out by the produced parton 
in the infinite momentum frame.

The limits of variation of $z_0$ are fixed by the exact kinematical relations 
$$
B - \sqrt{B^2 - \left( {Q_0 \over Q} \right)^2} \le z_0 \le B + 
\sqrt{B^2 - \left( {Q_0 \over Q} \right)^2}
\eqno(11)
$$
where $B = {1 \over 2} [1 + ({Q_0 \over Q})^2 - ({Q_1 \over Q})^2]$ is 
the scaled parton energy in the center of mass system 
and $\sqrt{B^2 - ({Q_0 \over Q})^2}$ its 
maximum scaled transverse momentum. In this way the scaled 
transverse momentum of the parton with energy fraction $z_0$ 
$$
{\vert \pt_0 \vert^2 \over Q^2} = \sqrt{B^2 - \left( {Q_0 \over Q} \right)^2} - 
(z_0 -B)^2 
\eqno(12)
$$
and its rapidity
$$
y_0 = y + {1 \over 2} \log { B + \vert z_0 - B \vert \over B - \vert z_0 - B \vert }
\eqno(13)
$$
are uniquely determined.
Rapidity of the second parton of virtuality $Q_1$ 
is obtained by energy-momentum conservation:
$$
y_1 = y + {\tanh}^{-1} \ \left[ {B \over B+1} {\tanh}\ |y_0-y| 
\right]
\eqno(14)
$$
Notice that only the first step has to be treated differently in 
rapidity because it corresponds to the degrading from the maximum 
allowed virtuality $W$ to the virtuality of the first splitting $Q$. In 
this case the rapidity of the ancestor is fixed by conservation laws and 
is given by 
$$
y = {\tanh}^{-1} \ \sqrt{1 - \left( {Q \over W} \right)^2} 
\eqno(15)
$$
The kinematical structure of the model is summarized in Figure 1.

The virtuality evolution equations in full phase space 
have been analytically solved in two cases: 
a Poissonian distribution has been obtained when only one produced 
parton can split 
and a geometric distribution is found when we neglect the 
virtuality conservation law in the production process.

We found that the numerical solution of the master equation for the MD in a single jet 
is well reproduced by a NB distribution in full phase space, in 
symmetric rapidity intervals and in $\pt$\ intervals. 
It is also found that the average number of clans approximately 
scales with jet energy in a fixed rapidity interval.

At this level of investigation, it should be clear that still open 
problems are the lack of the analytical 
solution of equation (6) and, in more general terms, 
the lack of a complete analytical 
 study of the parton evolution process in rapidity and $\pt$\ variables.

\beginsection III. The Generalized Simplified Parton Shower Model (GSPS) 

\vskip -0.6truecm 

In order to solve part of the problems indicated at the end of the previous 
section, we propose to incorporate in the SPS model the clan concept; 
we call this version of the model Generalized Simplified Parton Shower (GSPS) 
model. Accordingly, 
we decide now to pay attention for each event  to the ancestor which, 
splitting $n$ times, 
gives rise to $n$ subprocesses (one at each splitting) (see Figure 2) and 
we identify them with clans. Therefore in this model for 
a single event the concept 
of clan at parton level 
is no more as it was in \ref{1}, {\it i.e.}, only a 
{\it statistical} one. 
In the present picture the clans are independent active parton sources 
and {\it their number in each event coincides with the number 
of splittings of the ancestor, {\it i.e.}, with the number of steps in the 
cascade.} 

Notice that each clan generation is independent of previous history 
(it has no memory); thus the process is markoffian. Furthermore each 
generation process depends on the evolution variable only and is 
independent of the other variables of the process like the number of 
clans already present and their virtualities.
In the original version of SPS  the splitting function of 
the first step, $p(Q_0|Q)$, was different from the splitting 
function of all the other steps, $\beta(Q_0|Q)$, obtained by integrating 
the joint probability function $\P$
$$
\beta (Q_0 \vert Q) = \int_1^{Q-Q_0} dQ_1 \P 
\eqno(16)
$$
In order to generalize the SPS model as it stands 
we assume that 
virtuality conservation law is locally violated (although conserved globally) 
according to 
$$
1 \le Q_0 \le Q, \qquad 1 \le Q_1 \le Q 
\eqno(17)
$$
The upper limit of integration of (16) becomes 
$Q$, the normalization factor $K(Q)$ reduces to 1 and 
the process becomes homogeneous in the evolution variable since 
$$
\beta(Q_0|Q) = p(Q_0|Q) \int_1^Q p(Q_1|Q) dQ_1 = p(Q_0|Q)
\eqno(18)
$$
where $p(Q_0|Q)$ is factorized according to eq.~(7).
The approximation described by eq. (18), therefore,  
can be interpreted as the effect of local 
fluctuations in virtuality occurring at each clan emission.

This violation of the virtuality 
conservation law spoils of course the validity of the energy-momentum 
conservation law, which, in the SPS model, uniquely determines 
the rapidity 
of a produced parton, given its virtuality and the virtuality and 
rapidity of its germane parton (see eq. (14)). 
In the GSPS model 
the two produced partons at each splitting are independent 
both in virtuality and in  rapidity; however, 
the rapidity of each parton is bounded by the extension of phase space 
fixed by its virtuality and the virtuality of the parent parton:
$$
|y_i - y| \le \log {Q \over Q_i} 
\eqno(19)
$$
In conclusion, by weakening locally conservation laws, we 
decouple the production process of partons at each splitting. 
Consequently, the GSPS model allows to follow just a branch of the splitting, since 
each splitting can be seen here 
as the product of two independent parton emissions.
This consideration will be particularly useful in discussing the 
structure in rapidity of the model; in fact, it is implied that 
the Altarelli-Parisi kernel given in eq. 
(10) should be identified with
$$
p(y_0|Q_0Qy)  dy_0  p(y_1|Q_1Qy) dy_1 \propto 
{dz_0 \over z_0}  {dz_1 \over z_1} 
\eqno(20)
$$

In the following we propose to study clan formation in a parton shower 
first in virtuality variable and, next, in virtuality and rapidity 
variables.
The $\pt$\ dependence of clan properties in the GSPS model could be 
treated with the proper modifications in an analogous way.

In order to stress the fact that from now on we will pay attention to 
clan production, we use lowercase letters ($r$,$p$) when we refer to the 
probability of producing $N$ clans, whereas in section {\bf II} we used 
uppercase letters ($R$,$P$) for the probability of producing $n$ partons. 
In addition, in order to be extremely clear, we proceed to calculate 
first the probability for emitting $N$ clans without including the first 
step, $r_N$, and later the same probability by including it, $p_N$.

\subsection III.a. Clans formation in a parton shower: Full Phase Space

Let us consider the probability $r_N(Q_0|Q)$ 
that an ancestor parton, splitting at 
virtuality $Q \ge 2$ GeV, 
gets virtuality $Q_0$ after emitting $N-1$ clans. 

In the GSPS model this probability can be expressed 
in terms of the elementary splitting function, $p(Q_0|Q)$, 
(see Figure 3, solid line diagrams):
$$\eqalign{
r_0(Q_0|Q) =&~  0, \qquad r_1(Q_0|Q) = 0, \qquad 
r_2(Q_0|Q) = p(Q_0|Q), \cr
r_3(Q_0|Q) =&~ \int_{\max \{2,Q_0\}}^Q dQ_1 p(Q_0|Q_1) p(Q_1|Q), \cr
\noalign{\vdots}
r_N(Q_0|Q) =&~ \int_{\max \{2,Q_0 \}}^Q dQ_{N-2} 
\int_{\max \{2,Q_0 \}}^{Q_{N-2}} dQ_{N-3} 
\dots \cr
&~\int_{\max \{2,Q_0 \}}^{Q_2} dQ_1 
p(Q_0|Q_1) p(Q_1|Q_2) \dots p(Q_{N-2}|Q) \cr}
\eqno(21) 
$$
Notice that, since a parton of virtuality less than 2 GeV does not split, 
the probability $r_N(Q_0|Q)$ has a 
different form when the virtuality $Q_0$ is larger or smaller than  
2 GeV. 
By taking into account the 
factorization of $p(Q_0|Q)$ given in eq. (7), eq. (21) can be rewritten as:
$$\eqalign{
r_N&(Q_0|Q) = p(Q_0 \vert Q)  
\int_{\max \{2,Q_0 \}}^Q p(Q_{N-2}|Q_{N-2}) dQ_{N-2} \cr
&\int_{\max \{2,Q_0 \}}^{Q_{N-2}} p(Q_{N-3}|Q_{N-3}) dQ_{N-3} 
\dots 
\int_{\max \{2,Q_0 \}}^{Q_2} p(Q_1 \vert Q_1) dQ_1 
\cr}
\eqno(22)
$$
The multiple integral 
of the product of identical factors over the ordered domain 
${\max \{2,Q_0 \}} \le Q_1 \le Q_2 \le \dots \le Q_{N-2} \le Q$ 
can be solved 
by extending the integration domain to a (N-2)-dimensional hypercube 
of length $Q - {\max \{2,Q_0 \}}$. 
Then, by using the symmetry of the integrand, one gets
$$
r_N(Q_0|Q) = \cases{ p(Q_0|Q) {[\lambda(Q) - \lambda(Q_0) ]^{N-2} \over 
(N-2)!} &~ $Q_0 \ge 2$~GeV  \cr   
 p(Q_0|Q) {[\lambda(Q)]^{N-2} \over 
(N-2)!} &~ $Q_0 < 2$~GeV \cr}   
\eqno(23)
$$
with
$$
\lambda(u) \equiv \int_2^u p(u'|u')du' 
$$
The probability $r_N(Q)$ 
that a parton, splitting at virtuality $Q$, generates a shower with 
$N$ clans is given by integrating $Q_0$  
over the range allowed for a final parton: 
$$
r_N(Q) \equiv \int_1^2 dQ_0 r_N(Q_0|Q) 
\eqno(24)
$$
Since only the second of eq. (23) contributes, one obtains 
a shifted-Poisson distribution 
$$\eqalign{
r_0(Q) =&~ 0 \qquad \qquad r_1(Q) = 0 \cr 
r_N(Q) =&~ e^{-\lambda(Q)} {[\lambda(Q)]^{N-2} \over (N-2)!}  \qquad\qquad 
N \ge 2 \cr }
\eqno(25) 
$$
The average number of clans generated by an initial parton splitting at $Q$
follows:
$$
\bar N_r(Q) \equiv \sum_{N=0}^{\infty} N r_N(Q) =
\sum_{N=2}^{\infty} N e^{-\lambda(Q)} {[\lambda(Q)]^{N-2} \over (N-2)!} 
= \lambda(Q) + 2
\eqno(26)
$$
Now we can take into account the first step and study 
the multiplicity distribution for clans produced from 
an ancestor of maximum allowed virtuality $W$ (see Figure 4, solid line 
diagrams).
The probability that an ancestor  parton of maximum 
virtuality $W$ gets virtuality $Q_0$ after emitting $N-1$ clans, 
$p_N(Q_0|W)$, is given by:
$$\eqalign{
p_0(Q_0|W) =&~ 0, \qquad 
p_1(Q_0|W) = p(Q_0|W), \cr 
p_2(Q_0|W) =&~ \int_{\max \{2,Q_0\}}^W dQ p(Q_0|Q) p(Q|W), \cr 
\noalign{\vdots}
p_N(Q_0|W) =&~ \int_{\max \{2,Q_0 \}}^W dQ
\int_{\max \{2,Q_0 \}}^Q dQ_{N-2} \dots \cr
&\int_{\max \{2,Q_0 \}}^{Q_2} dQ_1 
p(Q_0|Q_1) p(Q_1|Q_2) \dots p(Q_{N-2}|Q)p(Q|W) \cr}
\eqno(27) 
$$
It should be pointed out that eq. (27) for $N \ge 2$ can be obtained also by rewriting 
$p_N(Q_0|W)$ in terms of $r_N(Q_0|Q)$:
$$
p_N(Q_0|W) = \int_{\max \{2,Q_0 \}}^W dQ  p(Q|W) r_N(Q_0|Q) 
\eqno(28)
$$
This relation for clans corresponds to eq. (3) for partons. 

Explicitly, 
$$
p_N(Q_0|W) = \cases{ p(Q_0|W) {[\lambda(W) - \lambda(Q_0) ]^{N-1} \over 
(N-1)!} &~ $Q_0 \ge 2$~GeV \cr   
p(Q_0|W) {[\lambda(W)]^{N-1} \over 
(N-1)!} &~ $Q_0 < 2$~GeV \cr}   
\eqno(29)
$$
Then, by integrating over $Q_0$ in the allowed virtuality range, 
one gets the multiplicity distribution  $p_N(W)$:
$$\eqalign{
p_0(W) =&~ 0  \cr
p_N(W) \equiv&~ \int_1^2 dQ_0 p_N(Q_0|W) 
= e^{-\lambda(W)} {[\lambda(W)]^{N-1} \over (N-1)!} \cr }
\eqno(30) 
$$
Being included in eq. (30) the graph corresponding to 
the case in which the ancestor does
not split (one-parton shower), 
$N$ starts from 1 and not from 2 as in eq. (26).

Eq. (30) is a shifted Poissonian distribution with average number of clans
given by:
$$
\bar N(W) \equiv \sum_{N=0}^{\infty} N p_N(W) =
\sum_{N=1}^{\infty} N e^{-\lambda(W)} {[\lambda(W)]^{N-1} \over (N-1)!} 
= \lambda(W) + 1
\eqno(31)
$$
which is fully consistent with eq. (26) via eq. (28).
We stress that this result has been obtained {\it a priori} in the 
present  
generalized version of the model, differently from 
what has been done previously in \ref{1} where the independent 
production of clans was an ``ansatz'' introduced {\it a posteriori} in order to explain 
the occurrence of NB regularity.

Eqs. (30) and (31) answer to our main 
questions in full phase space. In order to
introduce the subsequent discussion on rapidity dependence, we propose 
to study first the clan density in virtuality, 
{\it i.e.}, the number of clans emitted with virtuality $Q_0$ 
from an ancestor parton splitting at virtuality $Q$. 

We start by 
considering the probability that the $(N-1)^{\rm th}$ clan is emitted 
with virtuality $Q_0$. Now, it should be remarked
 that in the GSPS model the elementary splitting function for 
a parent parton of virtuality $Q$ to produce a clan of virtuality $Q_1$,
$p(Q_1|Q)$,  
has the same functional form of the probability to produce the 
associated parton of virtuality $Q_0$, $p(Q_0|Q)$. 
Therefore, the 
role played by the ancestor and the clan at the last step of the 
cascade can be interchanged; 
the probability that an ancestor, splitting at $Q$, gets 
virtuality $Q_0$ after emitting $N-1$ clans, $r_N(Q_0|Q)$, 
(see Figure 3, solid line diagrams and eqs. (21,23)) 
turns out to be equal to the probability to produce the 
$(N-1)^{\rm th}$ clan with virtuality $Q_0$ (see Figure 3, dashed line 
diagrams). 
Finally, by summing over $N$, the average number of clans 
produced with virtuality $Q_0$, $\rsigma(Q_0|Q)$, is obtained; 
the explicit 
expression of this probability is quite subtle, since one has to 
carefully take into account the role of the ancestor. 
For $Q_0 \ge 2$ GeV, the ancestor will generate other partons and 
will not contribute to the final clan density:
$$
\rsigma(Q_0|Q) \equiv \sum_{N=2}^{\infty} r_N(Q_0|Q) \qquad Q_0 \ge 
2~\rm GeV 
\eqno(32)
$$ 
For $Q_0 < 2$ GeV, the ancestor is considered as one parton clan and 
will contribute to $\rsigma(Q_0|Q)$ in addition to the term given in 
eq. (32); thus one has 
$$
\rsigma(Q_0|Q) \equiv \sum_{N=2}^{\infty} r_N(Q_0|Q) + 
r_{\rm anc}(Q_0|Q) \qquad Q_0 < 2~ \rm GeV
\eqno(33)
$$ 
where $r_{\rm anc}(Q_0|Q)$ is the probability to obtain an ancestor of 
virtuality $Q_0 < 2$ GeV in any number of steps and is given by 
$$
r_{\rm anc}(Q_0|Q) = \sum_{N=2}^{\infty} r_N(Q_0|Q) 
\eqno(34)
$$
Notice that to eq. (34) will contribute just the second of the two 
eqs. (23). Accordingly  
$$
r_{\rm anc}(Q_0|Q) = \sum_{N=2}^{\infty}  p(Q_0|Q) {[\lambda(Q)]^{N-2} \over 
(N-2)!} = p(Q_0|Q) e^{\lambda(Q)} 
\eqno(35) 
$$
and the full expression for the clan density in virtuality, 
$\rsigma(Q_0|Q)$, turns out to be 
$$
\rsigma(Q_0|Q) = \cases{  p(Q_0|Q) e^{\lambda(Q) - \lambda(Q_0)} 
&~ $Q_0 \ge 2$~ GeV \cr
 2 p(Q_0|Q) e^{\lambda(Q)} 
&~ $Q_0 < 2$~GeV \cr}
\eqno(36) 
$$
The consistency of this result can easily be checked: by integrating 
$\rsigma(Q_0|Q)$ over $Q_0$ in the full domain one should obtain 
indeed for 
the average number of clans, $\bar N_r(Q)$, the same expression which 
was calculated in eq. (26). In fact  
$$\eqalign{
\int_1^Q dQ_0 \rsigma(Q_0|Q) = &~
\int_1^2  dQ_0 2 p(Q_0|Q) e^{\lambda(Q)} + 
\int_2^Q  dQ_0 p(Q_0|Q) e^{\lambda(Q) - \lambda(Q_0)} = \cr 
=&~ 2 + \lambda(Q)  = \bar N_r(Q) \cr}
\eqno(37)
$$
The same procedure can be used for studying the 
density  of clan produced by an ancestor of maximum allowed virtuality 
$W$. 
We start by considering 
the probability to 
produce the $(N-1)^{\rm th}$ clan at virtuality $Q_0$ from the ancestor 
of maximum allowed virtuality $W$ (see Figure 4, dashed line diagrams). 
In the model this probability turns out to be 
equal to the probability that the ancestor of maximum allowed 
virtuality $W$ gets virtuality $Q_0$ after emitting $N-1$ clans, 
{\it i.e.}, to the probability $p_N(Q_0|W)$ given by eq. (29) (in  
Figure 4, solid line diagrams).

Notice that the showers with one parton only do not spoil our reasoning; 
however, they will contribute to the average number of clans 
produced with  
virtuality $Q_0$, $\psigma (Q_0|W)$, for $Q_0 < 2$ GeV. It follows 
$$
\psigma (Q_0|W) = \cases{ \sum_{N=2}^{\infty} p_N(Q_0|W) 
&~ $Q_0 \ge 2$~GeV \cr
\sum_{N=2}^{\infty} p_N(Q_0|W) + p_{\rm anc}(Q_0|W) 
 &~ $Q_0 < 2$~GeV \cr}
\eqno(38)
$$
where $p_{\rm anc}(Q_0|W)$ is given by definition by  
$$\eqalign{
p_{\rm anc}(Q_0|W) =&~ \sum_{N=1}^{\infty} p_N(Q_0|W) = 
\sum_{N=1}^{\infty} p(Q_0|W) {[\lambda(W)]^{N-1} \over (N-1)!} \cr
=&~ p(Q_0|W) e^{\lambda(W)} \cr}
\eqno(39)
$$
The sum starts of course from $N=1$ since, as we discussed previously, 
we must include the contribution of one-parton showers. 

Accordingly, one has:
$$
\psigma(Q_0|W) = \cases{ p(Q_0|W) \left[ e^{\lambda(W) - \lambda(Q_0)} - 1 
\right] &~ $Q_0 \ge 2$~GeV \cr
 p(Q_0|W) \left[ 2 e^{\lambda(W)} - 1\right] &~ $Q_0 < 2$~GeV \cr}
\eqno(40)
$$
The consistency of 
our calculation can be checked by integrating eq. (40) 
over $Q_0$ from $1$ to $W$; eq. (31) follows. 

Eq. (31) solves our first problem to determine analytically the average 
number of clans generated by an initial parton of maximum allowed 
virtuality $W$ in full phase space. It should be added 
that eq. (36) and eq. (40) are simply connected by the relation 
$$
\psigma (Q_0|W) =\cases{ \int_{Q_0}^W dQ p(Q|W) \rsigma(Q_0|Q) 
&~ $Q_0 \ge 2$~ GeV \cr
\int_2^W dQ p(Q|W) \rsigma(Q_0|Q) + p(Q_0|W) 
&~ $Q_0 < 2$~ GeV \cr}
\eqno(41)
$$
which is obtained by introducing eqs. (28), (32) and (39) into eq. (38).

In the following subsection, eq.~(41) will be extended to include rapidity
variable in order to calculate the average number of 
clans in symmetric rapidity intervals.

\subsection III.b. 
Clans formation in a parton shower: Symmetric rapidity intervals

In this subsection we extend 
our study to clan distributions in 
symmetric rapidity intervals $\Delta y = [-y_c,y_c]$.
We relate the probability  to have $N'$ clans 
in the symmetric rapidity interval $\Delta y$, 
$p_{N'}(y_c,W)$, to the corresponding probability 
 defined in full phase space (fps), 
$p_N(W)$, by using the following relation
$$
p_{N'}(y_c,W) = \sum_{N=N'}^{\infty} 
\Pi (N', y_c \vert N, {\rm fps}) p_N(W) 
\eqno(42)
$$
where $\Pi(N',y_c \vert N,{\rm fps})$ 
is the conditional probability to have $N'$ clans in $\Delta y$ 
when one has $N$ clans in full phase space; it contains 
all dynamical information on the {\it rapidity structure} 
of the production process. 

Following the discussion in subsection {\it III.a} and in particular the 
fact that {\it clans are not correlated in full phase 
space and the only allowed correlations are among particles within the 
same clan}, the conditional 
probability $\Pi(N',y_c \vert N,{\rm fps})$ turns out to be 
a positive binomial distribution 
$$
\Pi(N',y_c \vert N,{\rm fps}) = {N \choose N'} 
\pi^{N'} (1 - \pi)^{N-N'}
\eqno(43)
$$
where $\pi(y_c,W)$ is the probability to produce a single clan 
in the rapidity interval $\Delta y$ from the maximum allowed 
virtuality $W$. 
In terms of generating functions $f_{\rm clan}(z) \equiv \sum_{N=0}^{\infty} p_N z^N$, the 
above relation can be written as 
$$
f_{\rm clan}^{\rm \Delta y}(z) = f_{\rm clan}^{\rm fps}(\pi z + 1 - \pi)
\eqno(44)
$$
Being clans distributed in full phase space 
according to a shifted Poisson 
distribution (eq. (30)), {\it via} eq. (44), the probability 
generating function for clans in symmetric rapidity intervals 
has the following form 
$$
f_{\rm clan}^{\rm \Delta y}(z) = \bigl[ 
\pi(y_c,W) z + 1 - \pi(y_c,W) \bigr] 
e^{\lambda(y_c,W) (z-1)} 
\eqno(45) 
$$
where $\lambda(y_c,W) = \pi(y_c,W) \lambda(W)$.

Eq.\ (45), being the sum of two Poissonian distribution (the first a 
shifted one), has a nice physical meaning: the two terms correspond to the 
probability of having the ancestor within or outside the given rapidity 
interval.
Since in full phase space $\pi(y_c={\rm fps},W) = 1$, from eq. (45) one 
obtains the generating function for full phase space.
In addition, it should be pointed out 
that the void probability in a given 
rapidity interval, $p_0(y_c,W)$, is different from zero and is given 
by 
$$
p_0(y_c,W) = \left[ 1 - \pi(y_c,W) \right] e^{-\pi(y_c,W) \lambda(W)}
\eqno(46)
$$
When $y_c$ is very small, $\pi(y_c,W)$ 
tends to zero  and the unshifted Poissonian dominates. Thus, it can be 
stated that in the smallest rapidity intervals the clan multiplicity is 
 to a good approximation Poissonian and the full MD belongs 
to the class of Compound Poisson Distribution\ref{14}. 
$\pi(y_c,W)$ tends to 1 for $y_c$ at the border of phase space  and 
the exact full phase space shifted-Poisson distribution is approached.
Finally, when $\lambda(y_c,W)$ is sufficiently large, the 
shifted-Poissonian dominates at large $N$ (the tail of the distribution)
while the unshifted one dominates at small $N$ (the head of the 
distribution).
These facts might have some consequences in interpreting the anomalies 
found in NB behavior for small $N$ and the deviations from NB behavior 
in large rapidity intervals, which are controlled by the behavior of the 
distribution at large $N$.

From eq.~(45) it follows that the average number of clans is given by:
$$
\bar N(y_c,W) = \pi(y_c,W) \bar N(W)
\eqno(47)
$$

We are ready now to calculate the average number of clans in symmetric 
rapidity intervals.

The probability that the ancestor, splitting at 
virtuality $Q$ and rapidity $y$, gets virtuality $Q_0$ and rapidity $y_0
$ emitting $N-1$ clans, $r_N(Q_0y_0|Qy)$, 
can be expressed within the GSPS model (as for $r_N(Q_0|Q)$) 
in terms of elementary splitting functions according to the following 
formula (see Figure 3, solid line diagrams):
$$\eqalign{
r_N&(Q_0y_0|Qy) = \int_{\max \{2,Q_0 \}}^Q dQ_{N-2} \dots 
\int_{\max \{2,Q_0 \}}^{Q_2} dQ_1  
p(Q_0|Q_1) \dots p(Q_{N-2}|Q) \cr
&\int_{-\infty}^{\infty} dy_{N-2} \dots \int_{-\infty}^{\infty} dy_{1} 
Y(|y_0-y_1|,Q_0,Q_1) \dots  Y(|y_{N-2}-y|,Q_{N-2},Q) \cr}
\eqno(48) 
$$
where $Y(|y_i - y_{i+1}|,Q_i, Q_{i+1})$ is the probability that a parent 
parton with rapidity $y_{i+1}$ generates in a single
step a parton of rapidity $y_i$, being 
their virtualities $Q_{i+1}$ and $Q_i$ respectively. 

Inspired by the criterium of simplicity which we decided to follow from 
the beginning of our work and by the considerations discussed at the 
beginning of Section {\bf III}, we take for the probability 
$Y(|y_i - y_{i+1}|, Q_i, Q_{i+1})$ the simplified QCD-splitting 
kernel 
$$
Y(|y_i - y_{i+1}|, Q_i, Q_{i+1}) dy_i = 
P(z_i) dz_i \propto {dz_i \over z_i}
\eqno(49)
$$
By using eq. (13) properly rewritten, the kernel becomes
$$
Y(|y_i - y_{i+1}|, Q_i, Q_{i+1})  dy_i \propto dy_i 
\eqno(50)
$$
which after the normalization in the kinematically allowed domain
$|y_i - y_{i+1}| \le  \log (Q_{i+1}/Q_i)$ 
turns out to be:
$$
Y(|y_i - y_{i+1}|, Q_i, Q_{i+1})  dy_i = {dy_i \over 2 \log 
\left( {Q_{i+1} \over Q_i} \right) } 
\theta(\log (Q_{i+1}/Q_i) - |y_i - y_{i+1}|) 
\eqno(51)
$$
Notice that in the limit of $Q_{i+1} \to Q_i$ the function 
$Y(|y_i - y_{i+1}|, Q_i, Q_{i+1})$ reduces to a Dirac 
$\delta$-function; moreover, by inserting eq. (51) into eq. (48) and by 
integrating over $y_0$ in all the allowed domain one recovers 
immediately the full phase space result given in eq. (21).

In this equation, as well as in the case of full 
phase space, one can identify the
probability $r_N(Q_0y_0|Qy)$ with the probability that the $(N-1)^{\rm th}$ clan
is produced with virtuality $Q_0$ and rapidity $y_0$ from an ancestor which 
splits at virtuality $Q$ and rapidity $y$ (see Figure 3, dashed line 
diagrams). 

The integrations in the allowed domain in  rapidity of eq. (48) 
are awkward and quite difficult. The idea is to approximate the 
probability density given by eq. (51) with a gaussian of the following 
form:
$$
Y(|y_i - y_{i+1}|, Q_i, Q_{i+1})  = { 1 \over \sqrt{2 \pi} \sigma_i} 
\exp \left( - {|y_i - y_{i+1}|^2 \over 2 \sigma_i^2} \right)
\eqno(52)
$$
$\sigma_i^2$ being the width of the distribution given by 
$$
\sigma_i^2 \propto \log \left( {Q_{i+1} \over Q_i} \right)  = \alpha 
\log \left( {Q_{i+1} \over Q_i} \right) 
\eqno(53)
$$
This approximation corresponds to weaken locally 
energy-momentum conservation laws and simulates the onset of local 
fluctuations on the probability density defined by eq. (51). It is 
remarkable that with the just mentioned simplification 
the integrations on 
the rapidity domains can easily be performed and lead to the following 
result:
$$\eqalign{
\int_{-\infty}^{\infty} dy_{N-2} \dots &\int_{-\infty}^{\infty} dy_{1} 
Y(|y_0-y_1|,Q_0,Q_1) \dots  Y(|y_{N-2}-y|,Q_{N-2},Q) = \cr
=&~ {1 \over \sqrt{2 \pi} \left( \sum_{i=0}^{N-2} \sigma_i^2 \right)^{1/2}} 
\exp \left( - {|y_i - y_{i+1}|^2 \over 2 \sum_{i=0}^{N-2} 
\sigma_i^2} \right) \cr}
\eqno(54) 
$$
From eq. (53), one sees that the width of the right part of eq. (54) 
depends only on initial and final virtualities, {\it i.e.}, 
$$
\sum_{i=0}^{N-2} \sigma_i^2 = \alpha \sum_{i=0}^{N-2} (\log Q_{i+1} - 
\log Q_i ) = \alpha \log \left( {Q \over Q_0} \right) 
\eqno(55)
$$
It is clear that eq. (48) can be written now in terms of the probability 
$r_N(Q_0|Q)$ defined in eq. (21) as follows: 
$$\eqalign{
r_N(Q_0y_0|Qy) =&~  Y(|y_0 - y|, Q_0 Q) 
\int_{\max \{2,Q_0 \}}^Q dQ_{N-2} \dots \cr 
&\int_{\max \{2,Q_0 \}}^{Q_2} dQ_1  
p(Q_0|Q_1) \dots p(Q_{N-2}|Q) \cr 
 =&~  Y(|y_0 - y|, Q_0 Q) \times r_N(Q_0|Q) \cr}
\eqno(56) 
$$
By recalling eq. (23), one has:
$$
r_N(Q_0y_0|Qy) =\cases{ p(Q_0|Q) Y(|y_0 - y|, Q_0 Q)
{[\lambda(Q) - \lambda(Q_0) ]^{N-2} \over (N-2)!} &~ $Q_0 \ge 2$~  GeV 
\cr   
p(Q_0|Q) Y(|y_0 - y|, Q_0 Q) 
{[\lambda(Q)]^{N-2} \over (N-2)!} 
&~ $Q_0 < 2$~ GeV \cr}   
\eqno(57)
$$
In summary, the use of gaussian approximation for the step function 
(step function $\to$ gaussian) allows to see that the convolution of 
a generic number of distributions in rapidity domains given by the 
corresponding step functions is equivalent to a single gaussian 
distribution in rapidity. 
If we assume that 
this approximation can work the other way around (gaussian $\to$ step 
function) the initial form of the probability density 
$Y(|y_i - y_{i+1}|, Q_i, Q_{i+1})$ as given by eq. (51) is re-obtained 
restoring the correct kinematical domain for the total distribution. 

From eq. (57), it is quite easy to calculate the clan density 
in virtuality and rapidity 
produced by an ancestor which
splits at virtuality $Q$ and rapidity $y$. Once again, in this case 
the ancestor plays a very peculiar role: 
for $Q_0 \ge 2$  GeV  (the ancestor does not contribute) one has 
$$
\rsigma(Q_0y_0|Qy) \equiv \sum_{N=2}^{\infty} r_N(Q_0y_0|Qy) 
\qquad Q_0 \ge 2~  \rm GeV
\eqno(58)
$$ 
and for $Q_0 < 2$ GeV  (the ancestor contributes additively) 
the probability becomes
$$
\rsigma(Q_0y_0|Qy) \equiv
\sum_{N=2}^{\infty} r_N(Q_0y_0|Qy) + r_{\rm anc}(Q_0y_0|Qy) 
\eqno(59)
$$ 
where $r_{\rm anc}(Q_0y_0|Qy)$ is given by definition by  
$$
r_{\rm anc}(Q_0y_0|Qy) = \sum_{N=2}^{\infty} r_N(Q_0y_0|Qy)
\eqno(60)
$$
{\it i.e.}, 
$$
\rsigma(Q_0y_0|Qy) =
2 \sum_{N=2}^{\infty} r_N(Q_0y_0|Qy)
\qquad \qquad Q_0 < 2~ \rm GeV 
\eqno(61)
$$ 
Finally, we find 
$$
\rsigma(Q_0y_0|Qy) = \cases{  p(Q_0|Q) 
Y(|y_0 - y|, Q_0 Q) e^{\lambda(Q) - \lambda(Q_0)} 
&~ $Q_0 \ge 2$~ GeV  \cr
2 p(Q_0|Q) Y(|y_0 - y|, Q_0 Q) 
e^{\lambda(Q)} 
&~ $Q_0 < 2$~  GeV \cr}
\eqno(62) 
$$
Clan density in rapidity is calculated by 
integrating the above density over the virtuality $Q_0$ in the full 
virtuality range:
$$
\rsigma(y_0|Qy) \equiv \int_1^Q dQ_0 \rsigma(Q_0y_0|Qy)
\eqno(63)
$$
Let us study separately this integral for $Q_0 < 2$~ GeV and $Q_0 \ge 2$ 
GeV, {\it i.e.}, 
$$
\rsigma(y_0|Qy) \equiv r_{\Sigma, Q_0<2} + r_{\Sigma, Q_0\ge 2}
\eqno(64)
$$
For $Q_0< 2$ GeV one has
$$\eqalign{
r_{\Sigma, Q_0<2} =&~ 2 \int_1^2 dQ_0 p(Q_0|Q) 
Y(|y_0 - y|, Q_0 Q) e^{\lambda(Q)} = \cr
=&~ 2 e^{\lambda(Q)} \int_1^2 {dQ_0 \over Q_0} { A (\log Q_0)^{A-1} \over 
(\log Q)^A} {1 \over 2 \log \left( {Q \over Q_0} \right)} 
\theta \left(\log \left( {Q \over Q_0} \right) - | y_0 - y| \right) \cr}
\eqno(65)
$$
By defining then $ \log Q_0/ \log Q \equiv x$, 
equation (65) can be rewritten as
$$
r_{\Sigma, Q_0<2} = A { (\log Q)^{A-1} \over (\log 2)^A}  
\int_0^{\min (\log 2/\log Q, 1 - |y_0-y|/\log Q)}  
dx {x^{A-1} \over 1-x}  
\eqno(66)
$$
Notice that the integral in eq. (66) can be expressed as a series, {\it i.e.}:
$$
\int dx {x^{A-1} \over 1-x} = \sum_{n=0}^{\infty} {x^{A+n} \over 
A+n}
\eqno(67)
$$
As an example, let us consider the solution of eq. (66) corresponding to 
$A$=2. One gets:
$$\eqalign{
r_{\Sigma, Q_0<2} =&~ {2 \log Q \over [\log 2]^2} 
\biggl[ - 
{\log 2 \over \log Q} 
- \log \left( 1 - {\log 2 \over \log Q} \right) 
\biggr] \cr
&\hskip4.0truecm {\rm for} \ |y_0-y| \le \log Q - \log 2 \cr 
r_{\Sigma, Q_0<2} =& {2 \log Q \over [\log 2]^2} \left[ - 1 + 
{|y_0-y| \over \log Q} - \log \left( {|y_0-y| \over \log Q} \right) 
\right] \cr
&\hskip4.0truecm {\rm for} \ |y_0-y| > \log Q - \log 2 \cr}
\eqno(68)
$$
Coming now to the second term of eq. (64) ($Q_0 \ge 2$ GeV), one has
$$\eqalign{
r_{\Sigma, Q_0\ge 2} =&~ \int_2^Q dQ_0 p(Q_0|Q) 
Y(|y_0 - y|, Q_0 Q) e^{\lambda(Q) - \lambda(Q_0)}  = \cr
= \int_2^Q {dQ_0 \over Q_0} { A (\log Q_0)^{A-1} \over 
(\log Q)^A} &e^{\lambda(Q) - \lambda(Q_0)} 
{1 \over 2 \log \left( {Q \over Q_0} \right)} 
\theta \left(\log \left( {Q \over Q_0} \right) - | y_0 - y| \right) = \cr 
=&~ \int_2^Q { A dQ_0 \over Q_0 \log Q_0} 
{1 \over 2 \log \left( {Q \over Q_0} \right)} 
\theta \left(\log \left( {Q \over Q_0} \right) - | y_0 - y| \right) 
\cr }
\eqno(69)
$$
In terms of $x$, 
this equation can be rewritten as
$$
r_{\Sigma, Q_0\ge 2} = {A \over 2 \log Q} 
\int_{\log 2/\log Q}^{1-| y_0 - y|/\log Q}  dx  {1 \over x(1-x)}  
\eqno(70)
$$
Notice that the integral diverges for $x \to 1$, 
{\it i.e.}, for $|y_0-y| = 0$. However, this singularity is under 
control:
it is originated by our choice to follow just one branch of 
the splitting process, thus allowing $Q_0 \to Q$. Although the 
probability density is infinite for  $|y_0-y| = 0$, the resulting 
probability is finite for any finite range of $y_0$.
Therefore, one has 
$$
r_{\Sigma, Q_0\ge 2} =
{A \over 2 \log Q} \log \left[ \left({
\log Q - |y_0 - y| \over |y_0 - y|} \right) \left( 
{\log Q - \log 2 \over \log 2 } \right) \right], \qquad 
  |y_0-y| > 0 
\eqno(71)
$$
The expression of Eq.~(71) corresponding to $A=2$ is trivial. 

It should be noticed that Eq.~(71) is defined in the kinematically allowed
region 
$$
0 < |y_0 - y| \le \log Q - \log 2 
\; , 
\eqno(72)
$$
to be compared with the kinematically aloowed region of Eq.~(65) 
($|y_0 - y| \le \log Q$).

In conclusion, by adding eqs. (65) and (71) one obtains 
$\rsigma(y_0|Qy)$, the number of 
clans of rapidity $y_0$ generated 
from an ancestor splitting at virtuality $Q$ and 
rapidity $y$. 
The density of clans generated from an ancestor of maximum allowed 
virtuality $W$,  $\psigma (y_0|W)$, can be studied by integrating over 
$Q$ in the region shown in Figure 5 (see 
eq. (41) where analogous calculations are performed in 
full phase space). The solid line inside the $(y_0,Q)$-domain in Figure 
5 corresponds to eq. (15).
It follows
$$
\psigma (y_0|W) = \int_2^W dQ p(Q|W)  \rsigma(y_0|Qy) + 
p_{\rm anc}(y_0|W) 
\eqno(73)
$$
$p_{\rm anc}(y_0|W)$ being the probability that the ancestor does not 
split and has rapidity $y_0$ fixed by eq. (15):
$$
p_{\rm anc}(y_0|W) = \int_1^2 dQ_0 p(Q_0|W) 
\delta \left( y_0 - {\tanh^{-1}}\ \sqrt{1 - (Q_0/W)^2} \right)
\eqno(74)
$$
This contribution corresponds to the first diagram of Figure 4 and to 
the thick solid line in the lowest right corner of the 
$(y_0,Q)$ domain of Figure 5.
The $\delta$-function in eq. (74) can be rewritten in terms of $Q_0$ 
$$
\delta \left( y_0 - {\tanh^{-1}}\ \sqrt{1 - (Q_0/W)^2} \right) = 
\delta \left( Q_0 - {W \over {\cosh}\ y_0}  \right) {W {\sinh}\ y_0 
\over {\cosh}^2 \ y_0}  
\eqno(75)
$$
and eq. (74) becomes
$$\eqalign{
p_{\rm anc}&(y_0|W) =  
A\ {\tanh}\ y_0 { [ \log W - \log {\cosh}\ y_0 ]^{A-1} \over 
(\log W)^A} \cr 
&{\rm for} \  {\tanh^{-1}}\ \sqrt{1 - (2/W)^2}  \le y_0 \le 
{\tanh^{-1}}\ \sqrt{1 - (1/W)^2}  \cr}
\eqno(76)
$$
The density of clans in rapidity 
produced from an ancestor of maximum allowed virtuality $W$ can now be
calculated by using Eq.~(73).
Accordingly, the 
average number of clans in symmetric rapidity intervals is given by:
$$
\bar N(y_c, W) \equiv \int_{- y_c}^{y_c} dy_0 \psigma(y_0|W) 
\eqno(77)
$$
This result is remarkable. It should be pointed out that it depends on 
parameter $A$ only. 
Parameter $A$ controls -- as we have seen in \ref{13} -- 
the number of clans produced in full phase space, 
{\it i.e.}, the length of the cascade (since the number of clans in GSPS 
model is identified with the number of steps in the shower, more steps 
we have more extended turns out to be the shower). 
Analytical solutions of eq. (77) can be 
obtained explicitly for all integer positive values of parameter $A$, 
and numerical solutions can be computed for  all 
positive real values of $A$. 
We decided to postpone the discussion on the A-dependence of our formula 
to a forthcoming paper and to limit ourselves to indicate the structure 
of our formulae by solving analytically the case $A=2$. This limitation 
does not alter the main features of our results; in fact preliminary 
analysis shows that our results are qualitatively not altered by other 
choices of parameter $A$. 
In order to perform the integration, we used the approximation 
$$
{\tanh}^{-1}\ \sqrt{ 1 - \left( {Q \over W} \right)^2} \simeq \log W + 
\log 2 - \log Q 
\eqno(78)
$$
which slightly modifies the $(y_0,Q)$ domain (solid line in Figure 5) 
into the dashed lines in the same Figure.
We summarize in the Appendix the long and cumbersome calculations which 
are needed in order to integrate eq. (77). 

In Figure 6 
the clan density $\psigma(y_0|W)$ corresponding to eq. (73) is shown 
as a function of $y_0$ variable for maximum allowed virtualities 
$W = 50$ GeV, $W=$ 100 GeV and $W= 500$ GeV. 
The contribution of one-parton showers turns out to be 
negligible for this choice  of the $A$ value.
Notice that the height of the curve is 
decreasing and the width increasing with the energy. 
Convolution of clan density for two parton showers is shown in Figure 7. 
Notice that  the 
central dip at $y \simeq 0$ is slowly removed by increasing the energy of the 
initial parton. It should be kept in mind that the structure of Figures 6 
and 7 refers to clan production; found different behavior for 
parton production is not in contradiction with this behavior since we have 
still to include in our scheme parton production within a single clan.

In Figure 8 the average number of clans $\bar N(y_c,W)$ (eq. (77))  
is given as a 
function of rapidity width $y_c$ for the same $W$ values of 
Figure 6. 
Limitations on the rapidity intervals are determined by the 
available phase space corresponding to the different initial parton 
virtualities. 

Accordingly, 
the GSPS model predicts for the average number of clans $\bar N(y_c,W)$ 
at parton level in a single shower (jet):

\item{\it a)} 
a rising 
in rapidity width $y_c$ for 
different initial parton virtualities $W$ very close to linear 
for $1 < y_c < y_{\rm fps}$; 
 the rising is still linear but with a somewhat different slope for $y_c <1$.
Characteristic bending occurs finally for 
rapidity width $y_c \lsim y_{\rm fps}$;

\item{\it b)}
approximate (into 5\%) energy independence in a fixed rapidity interval 
$y_c$ for $W$ below 100 GeV. 
For higher virtualities deviations from energy independence become 
larger; they are into 20\% when comparing $\bar N(y_c,50\ {\rm GeV})$ 
and $\bar N(y_c,500\ {\rm GeV})$. It should be noticed that the average 
number of clans slowly decreases with virtuality;  
this behavior has been already observed in Monte Carlo simulations for 
single gluon jets\ref{8,16}.

In addition to the above results which are consistent with our 
expectations on clan properties in parton showers, the model shows energy 
independent behavior (see Figure 9) 
by normalizing the average number of 
clans produced in a fixed rapidity interval $|y| \le y_c$ to the 
corresponding average number in full phase space, and by expressing this 
ratio as a function of the rescaled rapidity variable $y_c^* \equiv y_c/
y_{\rm fps}$:
$$
\pi^*(y_c^*,W) \equiv {\bar N(y_c^* \  y_{\rm fps},W) \over \bar N(W)}
\eqno(79)
$$
This new regularity turns out to be stable for different choices of the 
parameter $A$. In Figure 9 a clean linear behavior is shown for the 
above ratio corresponding to the parameter value $A$=2. 

It should be pointed out that the asymptotic ($\log W \to \infty$) 
expression of the average number of clans for $A=2$ 
corresponding to the interval 
$\log 2 < y_c < \log W - 2\log 2$, which covers almost all the available phase
space,  is given by
$$\eqalign{
\bar N(y_c,W) \simeq&~ 2 y_c {2 \log 2 + \log (\log W/\log 2) 
\over \log W}  + \cr
+&~ {2 \over (\log W)^2} \biggl[  - {11 \over 3} y_c \log 2  + {4 \over 3} 
y_c (\log 2)^2  - 3 y_c \log 2  \log ( \log W) + \cr
+&~ y_c \log 2  \log (\log 2) - {1 \over 2} \left( y_c - \log 2 \right)^2 
\log (y_c - \log 2) + \cr 
+&~ {1 \over 2} \left( y_c + \log 2 \right)^2 \log (y_c + \log 2) \biggr] 
+ {\cal O} \left( {1 \over \log W} \right)^3 \cr}
\eqno(80)
$$
Eq. (80) can be obtained from eq. (A8) in the Appendix.
To first order this formula shows that asymptotically 
the average number of clans in a fixed rapidity interval grows linearly
 with $y_c$ and is slowly decreasing with the energy 
of the initial parton. By including second order contribution, $\bar N(
y_c,W)$ continues to be linear in $y_c$ and to decrease with the energy 
but the slope of the curve is different from that at first order; this 
difference is quite remarkable (a factor 1.5) at low energies (below 100 
GeV), but decreases at high energies.  
Notice that eq. (80) describes perfectly well the result for the average 
number of clans obtained by using the complete formula discussed in 
the Appendix in the same rapidity domain. 

The asymptotic expression ($\log W \to \infty$) 
of $\pi^*(y_c^*,W)$ is given by
$$\eqalign{
\pi^*(y_c^*,W) 
\simeq&~ {2 y_c^* \log \left( {\log W \over \log 2} \right) \over 
1 + 2 \log \left( {\log W \over \log 2} \right) } \cr 
 +&~ {[(y_c^*)^2-1] \log (1-y_c^*) - 
(1+y_c^*)^2 \log (1+y_c^*)  \over 
1 + 2 \log \left( {\log W \over \log 2} \right) }+ \cr 
+&~ {4 y_c^* \log 2 + (y_c^*)^2  \over 
1 + 2 \log \left( {\log W \over \log 2} \right) } + 
{\cal O}({1 \over \log W \log \log W}) \cr}
\eqno(81)
$$
Eq.~(81) describes into 2\% the result obtained by using the complete
formula after inserting eq.~(A8) into eq.~(79). It should be 
pointed out that this behavior is maintained even at higher energies, 
indicating a scaling behavior in a wide energy range ($50\div500000$ 
GeV). 
In the limit $W \to \infty$, eq. (81) becomes 
$$
\pi^*(y_c^*,W) \simeq y_c^* + {\cal O}({1 \over \log \log W})  
\eqno(82)
$$
It should be noticed that the slope of this curve is 10\% less than 
that predicted by eq. (81), 
showing that the asymptotic regime is reached very late.

\beginsection Conclusions

We have discussed a simplified model, based on essentials of QCD and 
local weakening of conservation laws for a shower (jet) originated by a 
parton of given maximum allowed virtuality and given rapidity. 
It generalizes previous results 
to the case in which  rapidity is added to virtuality evolution 
by assuming that 
clans are intermediate independent parton sources; clan production 
in this framework can be 
described as a Markoffian process. The new model, which we called 
Generalized Simplified Parton Shower model (GSPS), allows to determine 
clan multiplicity distribution in full phase space, 
which turns out to be a shifted 
Poissonian to be compared with the Poissonian behavior assumed in the 
standard interpretation of NB regularity. For clan multiplicity 
distributions in symmetric rapidity intervals, we obtain a combination 
of two Poissonian distributions. 
It is remarkable 
that the model predicts clan properties which show the same qualitative 
general trend observed both in experimental data and in Monte Carlo 
simulations.
In addition it has been shown that the ratio 
$\bar N(y_c^*\  y_{\rm fps},W)/\bar N(W)$ is linear and energy independent when 
plotted as a function of rescaled rapidity variable $y_c^*$.

In this paper attention has been payed to the average number of clans 
produced in a given rapidity interval by an initial parton of maximum allowed 
virtuality $W$, $\bar N(y_c,W)$; the next step in our program is to 
calculate the corresponding average number of partons per clan in the 
same rapidity intervals, $\bar n_c(y_c,W)$. We decided to postpone this 
calculation in the framework of the GSPS  to a forthcoming paper.

\beginsection Appendix

In this Appendix we present a short discussion of the 
calculation of the
integrals given in eqs.~(73) and (77) for $A=2$. 
Although the integration 
of the densities $\rsigma(y_0|Qy)$ and $\psigma(y_0|W)$ might be 
considered straightforward, it is indeed complicated by the appearance 
of very many terms. They correspond in case of $\rsigma(y_0|Qy)$ 
(eqs.~(68) and (71)) to four regions of phase space which, according to Figure 10, we 
label with Roman letters. They are 
$$
\eqalign{\strut [a]&\cr \strut [b]&\cr \strut [c]&\cr
         \strut [d]&\cr}\qquad
\eqalign{y-\log Q  <~ &y_0 <  y-\log Q+\log 2 \cr 
         y-\log Q+\log 2  <~ &y_0 <  y \cr
         y  <~ &y_0 <  y+\log Q-\log 2 \cr
         y+\log Q-\log 2  <~ &y_0 <  y+\log Q \cr}
\eqno(A1)
$$
The densities $\rsigma(y_0|Qy)$ are different in the regions [a],[b],
[c],[d], and given by:
$$\eqalignno{
\rsigma[a] =&~ {2 \over (\log W)^2}{2 \over (\log 2)^2}  
\biggl[ ( \log 2W - y_0) \log Q + (\log Q)^2 \log \log Q + &(A2) \cr 
-&~ 2 (\log Q)^2 - 
(\log Q)^2 \log ( \log 2W - y_0 -\log Q) \biggr]  \cr
\rsigma[b] =&~ {2 \over (\log W)^2} 
\biggl[ -  {2\log Q \over \log 2} - {2 (\log Q)^2 \log ( \log Q - \log 2) \over 
(\log 2)^2} +\cr 
+&~ {2 (\log Q)^2 \log \log Q \over (\log 2)^2}    
+ \log(2\log Q + y_0 - \log 2W) + &(A3)\cr 
-&~ \log (\log 2W - y_0 - \log Q)  
+\log (\log Q - \log 2) - \log \log 2 \biggr] \cr 
\rsigma[c] =&~ {2 \over (\log W)^2} 
\biggl[ -  {2 \log Q \over \log 2} - {2 (\log Q)^2 \log ( \log Q - \log 2) \over 
(\log 2)^2} + \cr 
+&~ {2 (\log Q)^2 \log \log Q \over (\log 2)^2}  
+ \log (\log 2W - y_0) + &(A4)\cr 
-&~ \log (y_0 - \log 2W + \log Q) 
+ \log (\log Q - \log 2) - \log \log 2 \biggr] \cr 
\rsigma[d] =&~ {2 \over (\log W)^2}{2 \over (\log 2)^2} 
\biggl[ (y_0 - \log 2W) \log Q + &(A5) \cr 
-&~ (\log Q)^2 \log (y_0 -  \log 2W +\log Q) + (\log Q)^2 \log \log Q 
\biggr] \cr}
$$
where the approximation defined in eq. (78) for $y$ has been used. 

Accordingly, the density $\psigma(y_0|W)$ of eq.~(73) results to be 
different in the above mentioned regions. Its integration over $y_0$ 
variable leads to divide the integration domain into five regions which we 
label by Greek letters in Figure 10, {\it i.e.}:
$$
\eqalign{\strut (\alpha)&\cr \strut (\beta)&\cr \strut (\gamma)&\cr
         \strut (\delta)&\cr \strut (\epsilon)&\cr}\qquad
\eqalign{  -\log W+\log2  <~ &y_0 <  -\log W+2\log 2\cr
         -\log W+2\log 2  <~ &y_0 <  \log 2 \cr
                  \log 2  <~ &y_0 <  \log W-\log 2\cr
           \log W-\log 2  <~ &y_0 <  \log W\cr
                  \log W  <~ &y_0 <  \log W+\log 2\cr}
\eqno(A6)
$$
The analytic solution  of the integral in eq.~(77) turns out to be 
different for $y_c$ in the following different domains:
$$
\eqalign{            &y_c <  \log 2\cr
          \log 2  <~ &y_c <  \log W - 2 \log 2 \cr 
\log W - 2 \log 2  <~ &y_c <  \log W - \log 2 \cr
\log W -  \log 2  <~ &y_c <  \log W  \cr
\log W   <~ &y_c <  \log W + \log 2 \cr}
\eqno(A7)
$$
In order to give the flavor of the cumbersome and long calculations 
which lead to the final form of $\bar N(y_c,W)$, whose asymptotic 
expression has been given in eq.~(80), we show the result 
for $y_c$ in the domain 
$\log 2 < y_c < \log W - 2\log 2$, {\it i.e.}:
\vfill\eject
$$
\eqalign{ \bar N(&\yc,W) =
{2\over {{(\lw)^2}}}
\left\{ {{\lw\,\la}\over 2} + {{{(\la)^2}}\over 2} +
       {{5\,\yc,\lw}\over 3} -
       {{{\yc\,(\lw)^2}}\over {3\,\la}} +           \right. \cr\spac &
       {{7\,\yc\,\la}\over 6} + 2\,\yc\,\lw\,\la +    
       {{4\,{\yc\,(\la)^2}}\over 3} + {{{\yc^2}}\over 2} +
       {{{\yc^3}}\over {3\,\la}} +                    \cr\spac &
       \left( {(\lw)^2} + {{4\,\yc\,{(\lw)^3}}\over
            {3\,{(\la)^2}}} \right) \,\log (\lw) +      \cr\spac &
       \left( 2\,\yc\,\lw -
          {{4\,\yc\,{(\lw)^3}}\over {3\,{(\la)^2}}} -
          {{2\,\yc\,\la}\over 3} \right) \,
        \log (\lw - \la) +                            \cr\spac &
       \left( -\lw + \la \right) \,\yc\,\log (\la) +    \cr\spac &
       {{{\left( -\lw + \yc \right) }^2}\over {6\,{(\la)^2}}}\,
           \left( -{(\lw)^2} - 4\,\lw\,\la -
             3\,{(\la)^2} + 2\,\yc\,\lw+          \right. \cr\spac & \left.
             4\,\yc\,\la - {\yc^2} \right) \,
           \log (\lw - \yc) -                      \cr\spac &
       {{{{\left( -\lw + \la + \yc \right) }^2}\,
           \log (\lw - \la - \yc)}\over 4} +         \cr\spac &
       {\log (\lw + \la - \yc) \over 12\,(\la)^2}
             \left[ 2\,{(\lw)^4} + 8\,{(\lw)^3}\,\la +
             3\,{(\lw)^2}\,{(\la)^2} +              \right. \cr\spac &
             2\,\lw\,{(\la)^3} + 5\,{(\la)^4} -     
             8\,\yc\,{(\lw)^3} -
             24\,\yc\,{(\lw)^2}\,\la -         \cr\spac &
             18\,\yc\,\lw\,{(\la)^2} -         
             14\,\yc\,{(\la)^3} +              
             12\,{(\lw)^2}\,{\yc^2} +          
             24\,{\yc^2}\,\lw\,\la +         \cr\spac & \left.
             15\,{\yc^2}\,{(\la)^2} -
             8\,{\yc^3}\,\lw - 8\,{\yc^3}\,\la +
             2\,{\yc^4} \right] +             \cr\spac &
       {\log (\lw + \yc)\over 6(\la)^2}\,
           \left( {(\lw)^2} + 4\,\lw\,\la +
             3\,{(\la)^2} + 2\,\yc\,\lw +          \right. \cr\spac &\left.
             4\,\yc\,\la + {\yc^2} \right) \,
            {\left( \lw + \yc \right) }^2 -
       {{{\left( -\la + \yc \right) ^2}\,
           \log (-\la + \yc)}\over 2} +             \cr\spac &
       {{{{\left( \la + \yc \right) }^2}\,
           \log (\la + \yc)}\over 2} -              \cr\spac & \left.
       {{{{\left( \lw + \la + \yc \right) }^4}\,
           \log (\lw + \la + \yc)}\over {6\,{(\la)^2}}} \right\}
}
\eqno(A8)
$$

\vfill\eject
\beginsection References

\item{[1]}
A. Giovannini and L. Van Hove, Z.\ Phys.\ C30 (1986) 391

\item{[2]}
A. Giovannini and L. Van Hove, Acta Phys.\ Pol.\ B19 (1988) 495

\item{[3]}
P.\ Abreu et al., DELPHI Collaboration, Z.\ Phys.\ C56 (1992) 63

\item{[4]}
M.\ Arneodo et al., EMC Collaboration, Z.\ Phys.\ C35 (1987) 335

\item{[5]}
G.J.\ Alner et al., UA5 Collaboration, Phys.\ Lett.\ B160 (1985) 193

\item{[6]}
L. Van Hove and A. Giovannini, Acta Phys.\ Pol.\ B19 (1988) 917

\item{[7]}
A.\ Giovannini, S.\ Lupia and R.\ Ugoccioni, 
Nucl.\ Phys.\ B (Proc.\ Suppl.) 25B (1992) 115

\item{[8]}
F. Bianchi, A. Giovannini, S. Lupia and R. Ugoccioni,
Z. Phys.\ C58 (1993) 71;

\item{}
F. Bianchi, A. Giovannini, S. Lupia and R. Ugoccioni,
in Proceedings of the
XXII International Symposium on Multiparticle Dynamics,
(Santiago de Compostela, Spain, 1992), World Scientific, Singapore,
1993, p.~143

\item{[9]}
G. Gustafson, Nucl. Phys. B392 (1993) 251;  

\item{}
G. Gustafson and M. Olsson, Nucl. Phys. B406 (1993) 293 

\item{[10]}
W. Ochs, Z. Phys. C -- Particles and Fields 23 (1984) 131;

\item{}
W. Ochs, J. Wosiek, Phys. Lett. B304 (1993) 144

\item{[11]}
I.M. Dremin, R.C. Hwa, ``Quark and gluon jets in QCD: factorial and 
cumulants moments'', preprint OITS 531, December 1993

\item{[12]}
Yu. L. Dokshitzer, Phys. Lett. B305 (1993) 295

\item{[13]}
R.\ Ugoccioni and A.\ Giovannini, 
Z.\ Phys.\ C53 (1992) 239;

\item{}
A.\ Giovannini and R.\ Ugoccioni, invited talk, 
in: Fluctuations and Fractal Structure, 
Proceedings of the Ringberg Workshop on Multiparticle
Production (Ringberg Castle, Germany, 1991), eds.~R.C.~Hwa, 
W.~Ochs and N.~Schmitz, World Scientific,
Singapore, 1992, p.~264

\item{[14]}
S. Lupia, A. Giovannini and R. Ugoccioni,
Z. Phys.\ C59 (1993) 427

\item{[15]}
Yu. L. Dokshitzer, V.A. Khoze, A.H. Mueller and S.I.Troyan, {\it Basics 
of Perturbative QCD}, Editions Fronti\`eres, Gif-Sur-Yvette, 1991

\item{[16]}
R. Ugoccioni, A. Giovannini and S. Lupia,
``The Generalized Simplified Parton Shower Model'', DFTT~52/93,
to be published in the Proceedings of the XXIII International Symposium
on Multiparticle Dynamics, Aspen, CO, USA, 12-17 September 1993

\vfill\eject
\beginsection Figure Captions

\noindent
{\bf Fig. 1.}\ The structure of the Simplified Parton Shower 
model (SPS) for parton production in virtuality $Q$ and rapidity $y$. 
The degrading from the maximum allowed virtuality $W$ is given by the 
product of eqs.~(7) and (15). The splitting at $(Q,y)$ is controlled by 
the product of eqs.~(4), (10) and (14).  
Each dot represents a parton which further generates.
Notice that here $Q_0 + Q_1 \le Q, \ 
|y_0 - y| \le \log {Q \over Q_0}$.

\noindent
{\bf Fig. 2.}\ The structure of the Generalized Simplified Parton Shower 
model (GSPS) for clan production in virtuality $Q$ and rapidity $y$. 
The production process is decoupled at each splitting both in virtuality 
and in rapidity (see eqs.~(18) and (20)).  Each blob represents a clan. 
Notice that here 
$Q_0 \le Q, \   Q_1 \le Q, \ 
|y_0 - y| \le \log {Q \over Q_0}, \ 
|y_1 - y| \le \log {Q \over Q_1}$.

\noindent
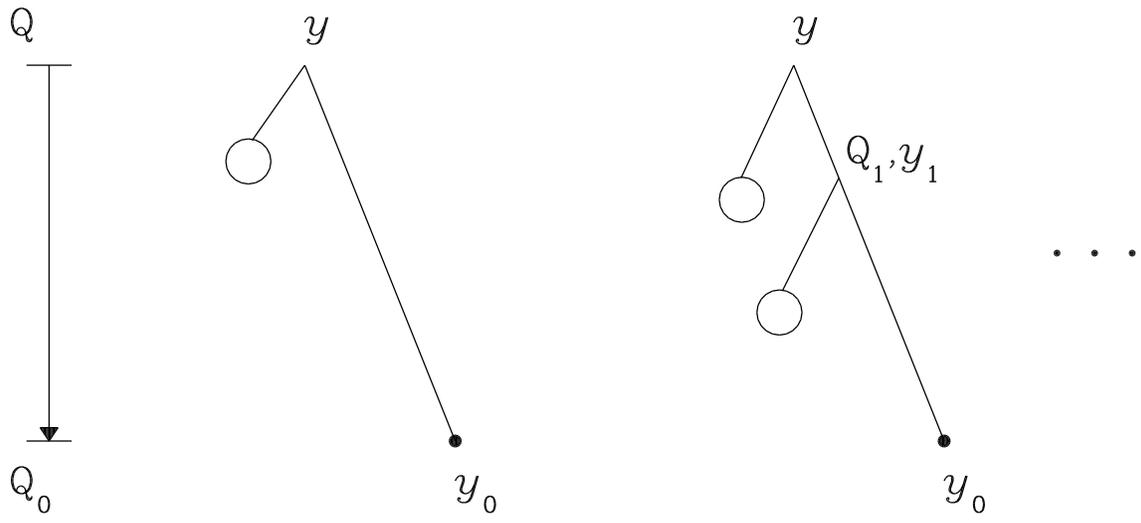
{\bf Fig. 3.}\ Diagrams for the probability that an ancestor, 
splitting at virtuality $Q$ and rapidity $y$ gets virtuality $Q_0$ and 
rapidity $y_0$ after emitting $N-1$ clans, 
$r_N(Q_0y_0|Qy)$ (solid line); dashed diagrams 
correspond to the interchange of the final 
ancestor and the last produced clan, {\it i.e.}, to the probability to 
produce the $(N-1)^{\rm th}$ clan with virtuality $Q_0$ and rapidity 
$y_0$ from an ancestor splitting at virtuality $Q$ and rapidity $y$, 
which can be described by the same $r_N(Q_0y_0|Qy)$.

\noindent
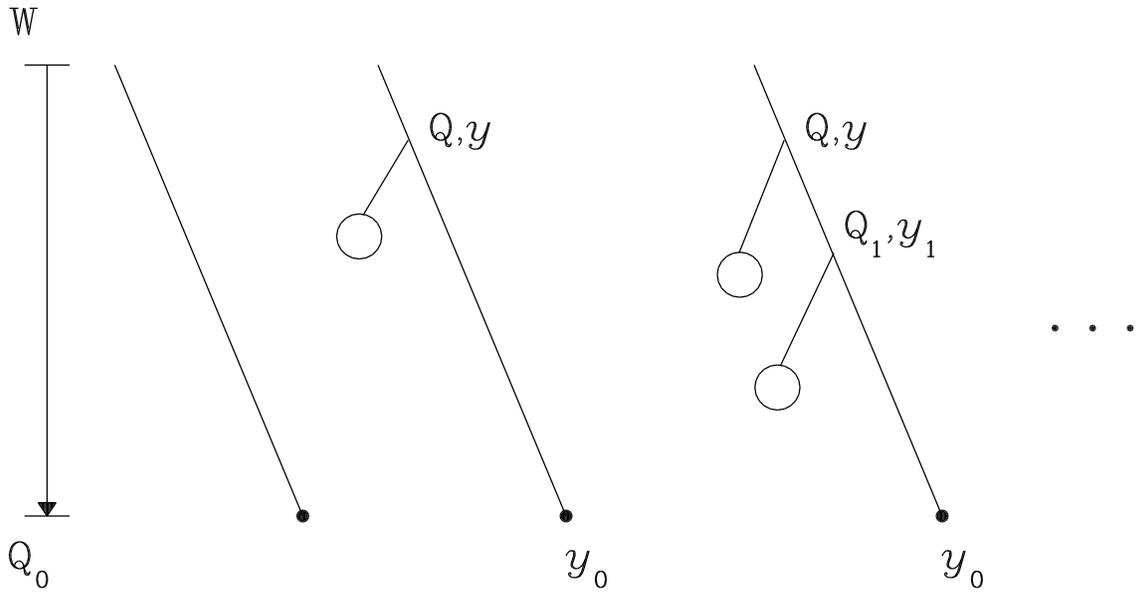
{\bf Fig. 4.}\ Diagrams for the probability that an ancestor of 
maximum allowed virtuality $W$ gets virtuality $Q_0$ and 
rapidity $y_0$ after emitting $N-1$ clans, 
$p_N(Q_0y_0|W)$ (solid line); dashed diagrams 
correspond to the interchange of the final 
ancestor and the last produced clan, {\it i.e.}, to the probability to 
produce the $(N-1)^{\rm th}$ clan with virtuality $Q_0$ and rapidity 
$y_0$ from an ancestor of maximum allowed virtuality $W$, 
which can be described by the same $p_N(Q_0y_0|W)$.

\noindent
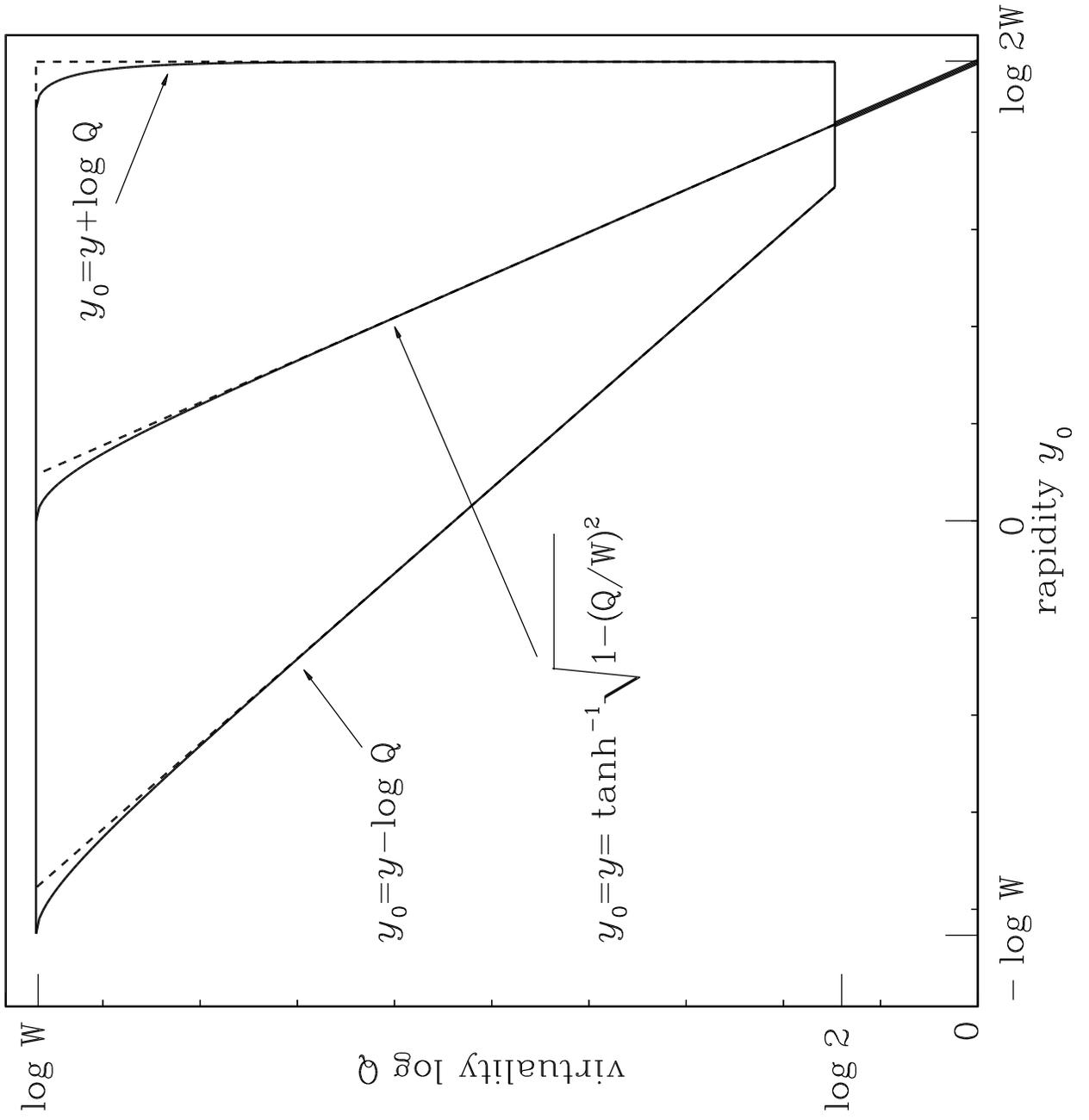
{\bf Fig. 5.}\ Phase space domain in virtuality and rapidity 
for an ancestor parton of maximum allowed virtuality $W$. 
Dashed lines represent the approximation used for explicit 
calculations (eq.~(78)). 
The thick solid line in the lowest right corner  corresponds 
to the contribution of one-parton shower, {\it i.e.}, 
to the first diagram in Figure 4. In the central part of the figure we 
show the line corresponding to eq.~(15). 

\noindent
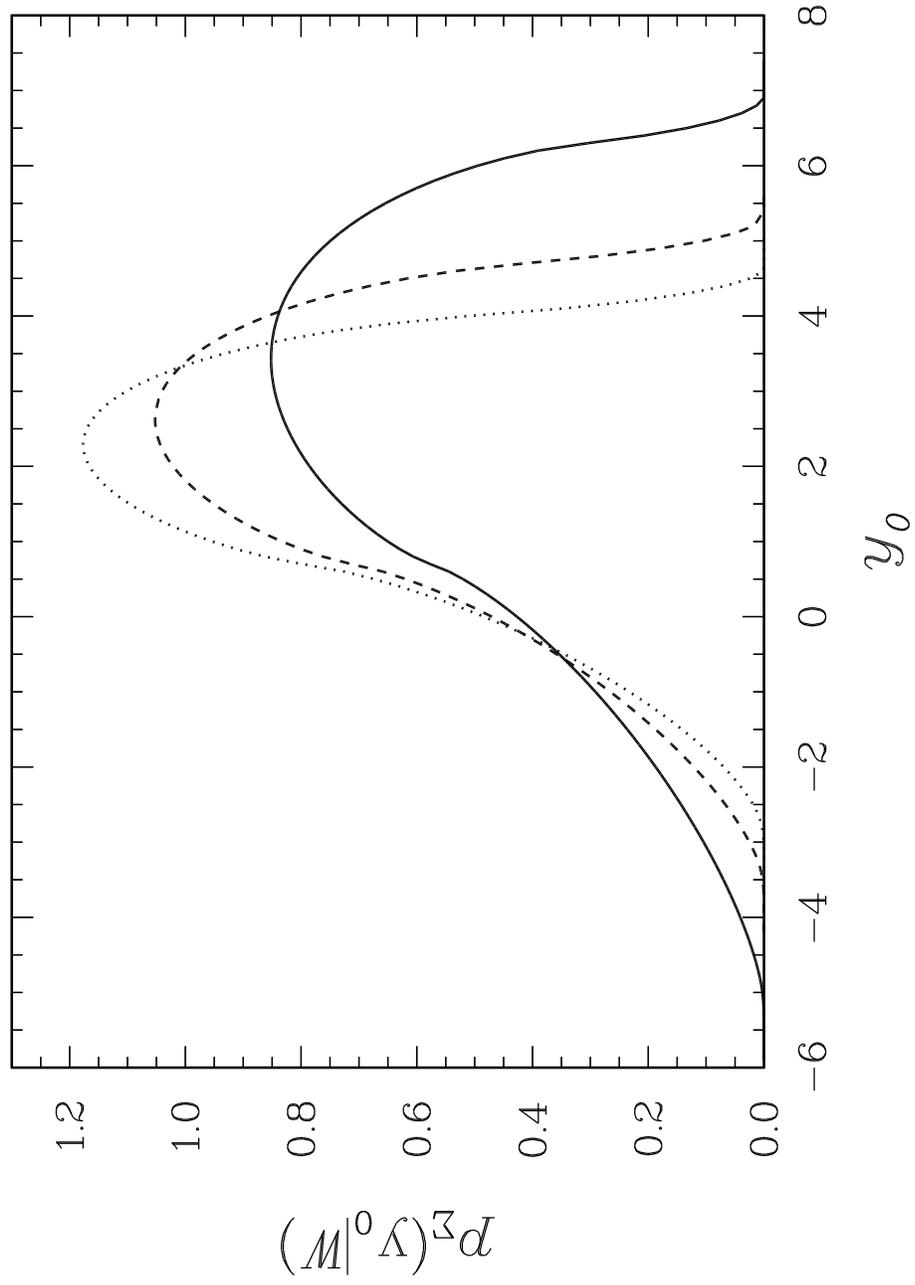
{\bf Fig. 6.}\ Clan density $\psigma(y_0|W)$ for one shower 
at $W$ = 50 GeV (dotted line), 100 GeV (dashed line) and 500 GeV (solid 
line). 

\noindent
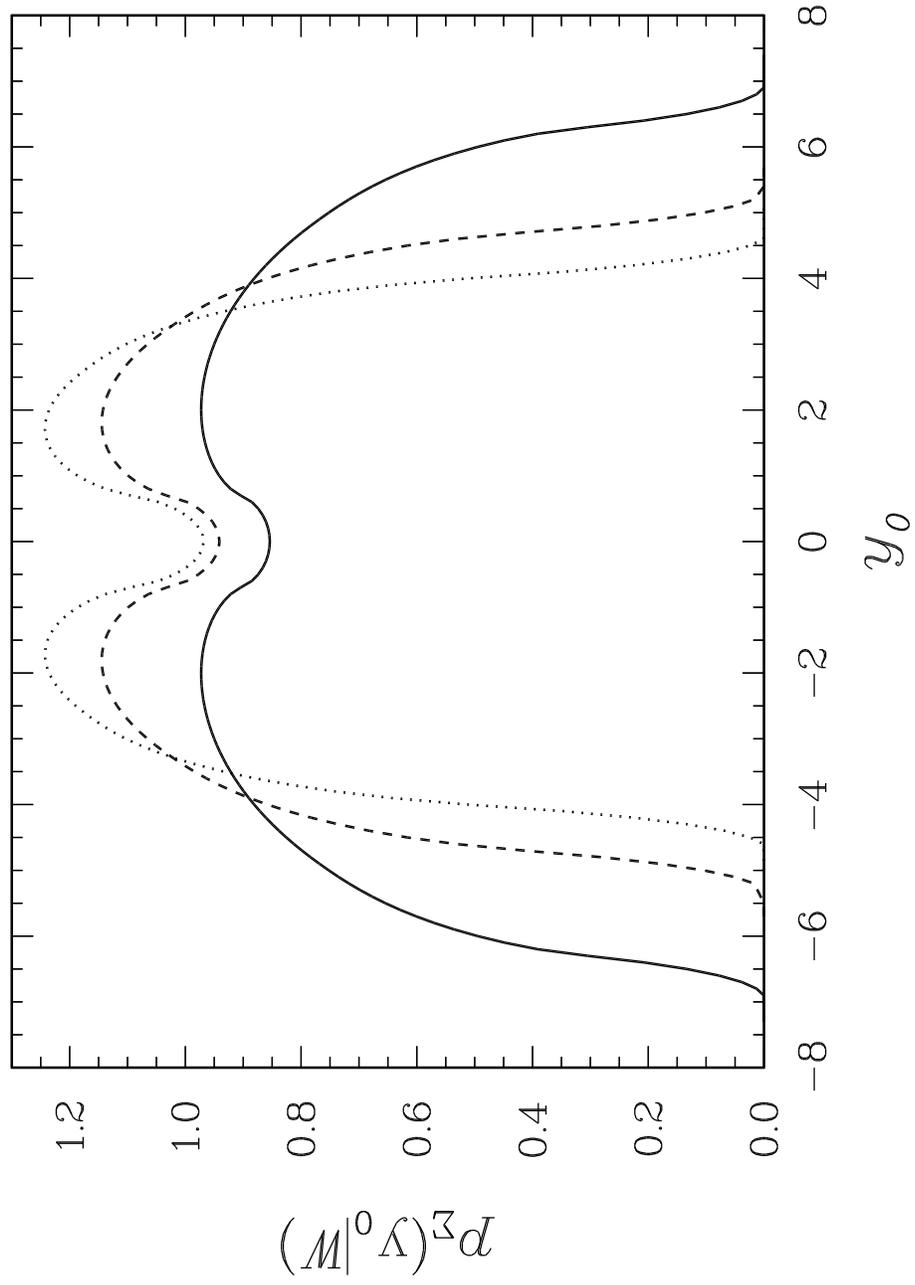
{\bf Fig. 7.}\ Clan density $\psigma(y_0|W)$ resulting by the 
addition of two back-to-back showers 
at $W$ = 50 GeV (dotted line), 100 GeV (dashed line) and 500 GeV (solid 
line). 

\noindent
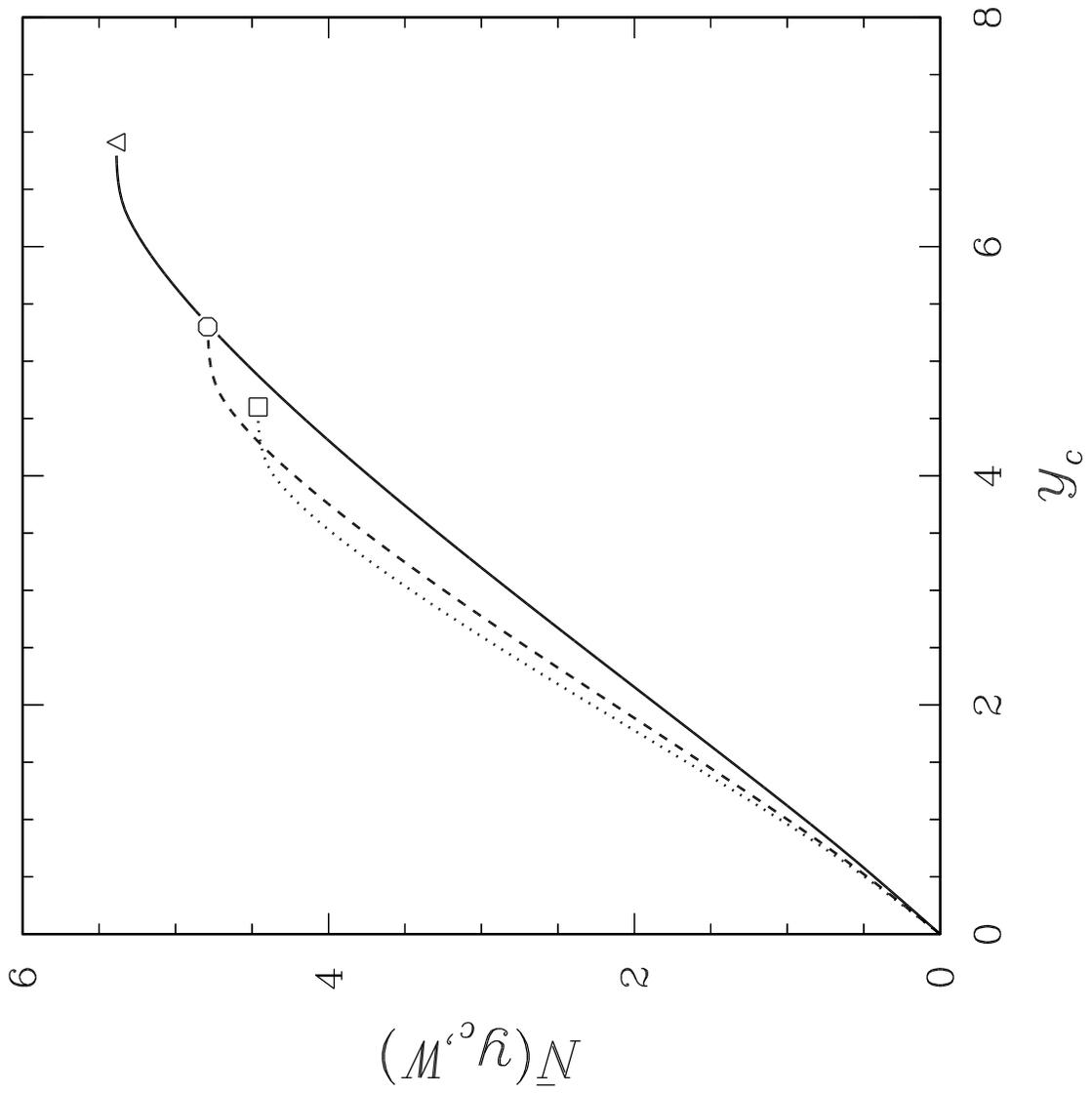
{\bf Fig. 8.}\ Average number of clans $\bar N(y_c,W)$ for a 
shower as a function of rapidity width $y_c$  
at $W$ = 50 GeV (dotted line), 100 GeV (dashed line) and 500 GeV (solid 
line). The symbols represent the values of the average number of clans 
in full phase space. 

\noindent
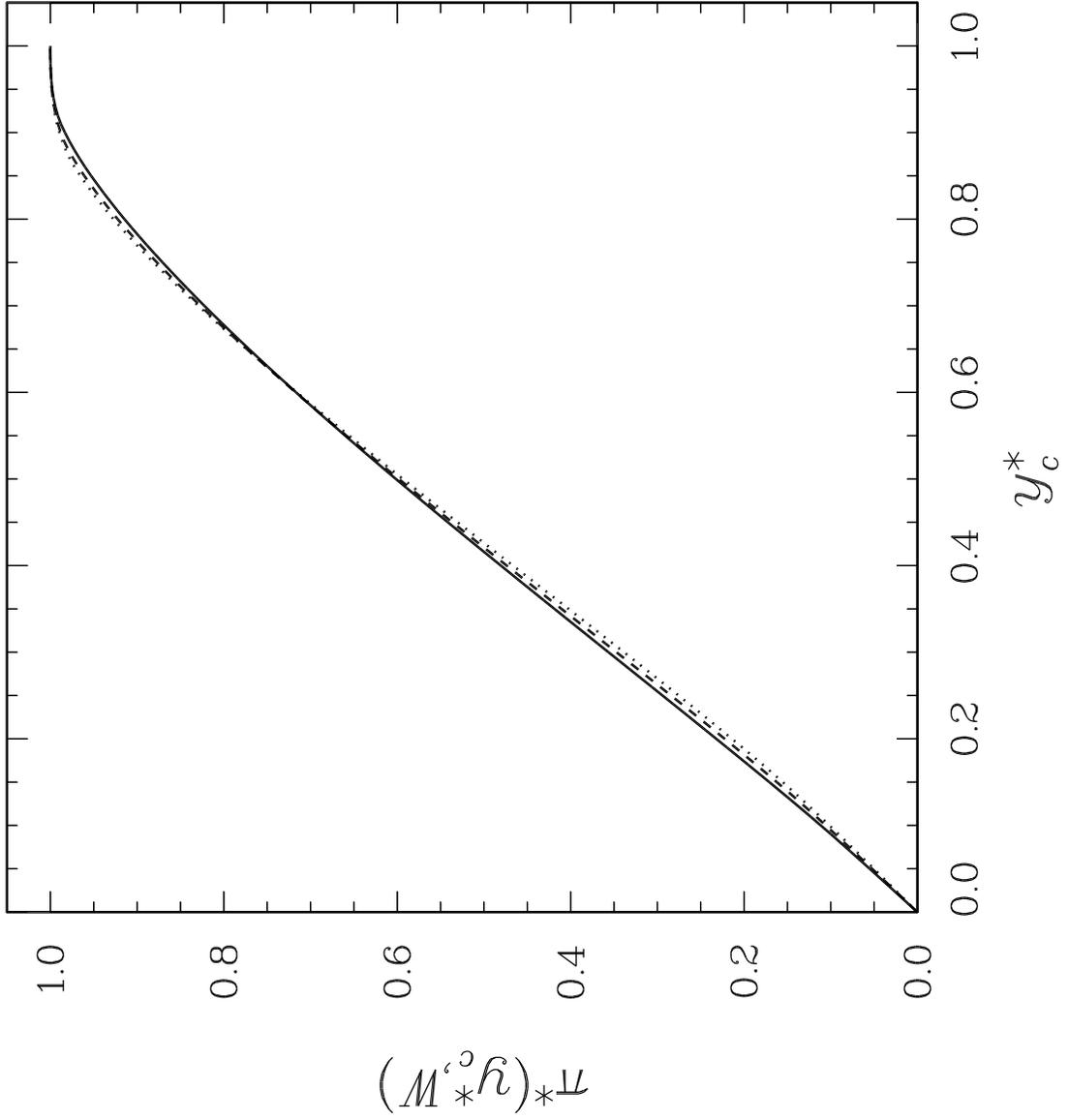
{\bf Fig. 9.}\  Normalized average number of clan 
$\pi^*(y_c^*,W)$ for a shower 
as a function of rescaled rapidity
$y^*_c = y_c/y_{\rm fps}$ 
at $W$ = 50 GeV (dotted line), 100 GeV (dashed line) and 500 GeV (solid 
line). 
 
\noindent
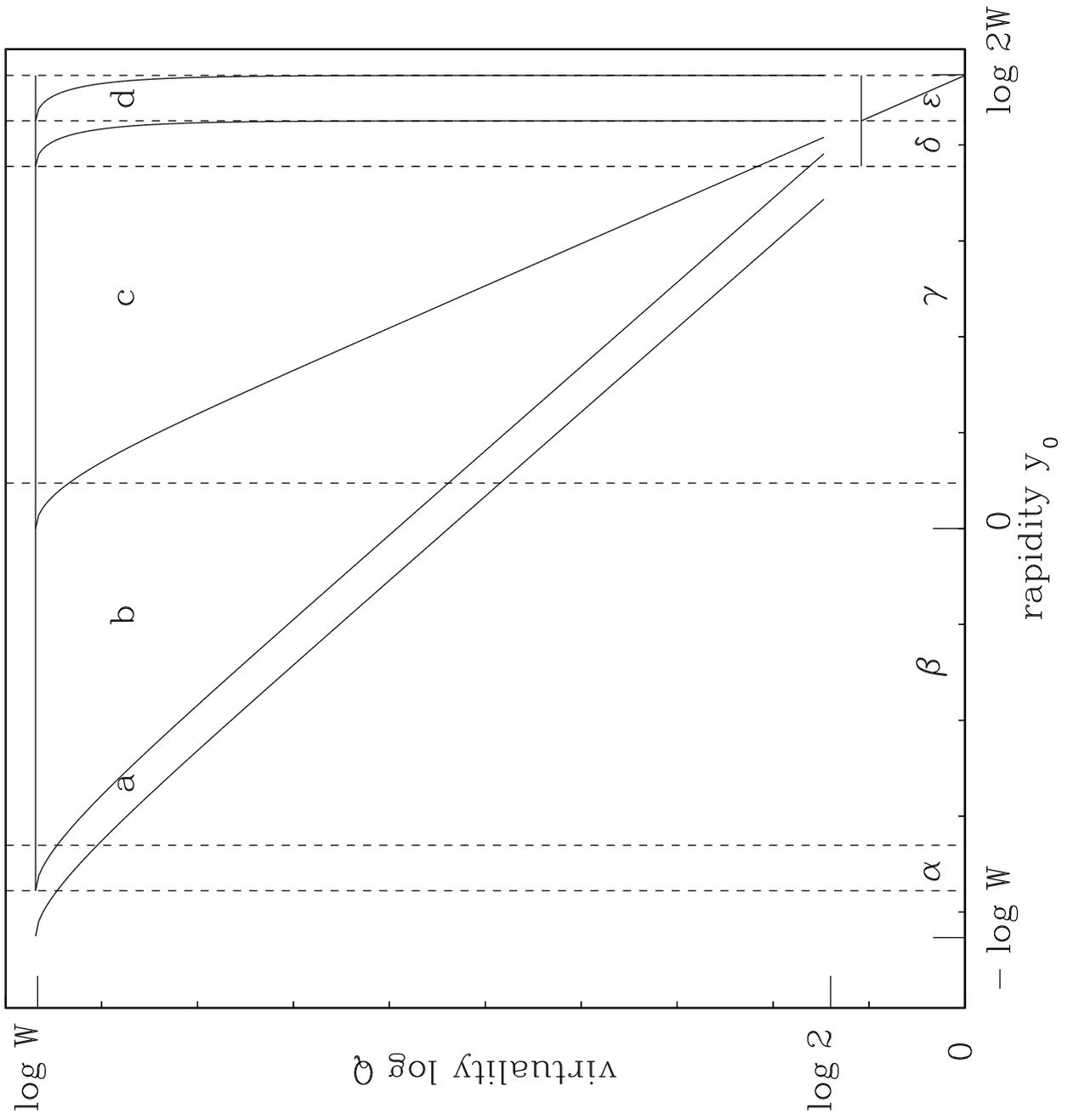
{\bf Fig. 10.}\ Phase space domain in virtuality and rapidity 
for an ancestor parton of maximum allowed virtuality $W$. 
Different domains defined in eq.~(A1) are labelled by Roman letters; 
different integration domains of eq.~(A6) are labelled by Greek letters.

\bye